\documentclass[%
 aip,
 jmp,%
 amsmath,amssymb,
 reprint,%
]{revtex4-1}

\usepackage{subfigure}
\usepackage{graphicx}
\graphicspath{{./}{./images/}{./images_all/}}
\DeclareGraphicsExtensions{.eps}

\usepackage{dcolumn}
\usepackage{bm}

\begin{document}

\preprint{AIP/123-QED}

\title[APPLIED PHYSICS LETTERS xx (2010)]{Energy-efficient mixed mode switching of a multiferroic nanomagnet for logic and memory}

\author{Kuntal Roy}
\email{royk@vcu.edu.}
\author{Supriyo Bandyopadhyay}
\affiliation{Department of Electrical and Computer Engineering, Virginia Commonwealth University, Richmond, VA 23284, USA}
\author{Jayasimha Atulasimha}
\affiliation{Department of Mechanical and Nuclear Engineering, Virginia Commonwealth University, Richmond, VA 23284, USA}

\date{\today}

\begin{abstract}
In magnetic memory and logic devices, a magnet's magnetization is usually flipped with a spin polarized current delivering a spin transfer torque (STT). This mode of switching consumes too much energy and considerable energy saving can accrue from using a {\it multiferroic} nanomagnet switched with a combination of STT and mechanical stress generated with a voltage (VGS). The VGS mode consumes less energy than STT, but cannot rotate magnetization by more than 90$^{\circ}$, so that a combination of the two modes is needed for energy-efficient switching. \end{abstract}

\maketitle


In magnetic logic and memory devices, the magnetization of a nanomagnet is usually switched with a spin-polarized current delivering a spin transfer torque~\cite{RefWorks:7,RefWorks:124}. This is a well-established technique that has been experimentally demonstrated in different systems~\cite{RefWorks:43,RefWorks:32} and is widely used for spin transfer torque random access memory (STTRAM).

Unfortunately, this method of switching dissipates too much energy and a more energy-saving approach is to rotate the magnetization of a {\it multiferroic} nanomagnet -- consisting of a piezoelectric layer and a magnetostrictive layer -- with an electrostatic potential applied to the piezoelectric layer~\cite{RefWorks:154}. The applied voltage generates stress in the magnetostrictive layer, which rotates its magnetization. Such rotations have been experimentally demonstrated~\cite{ RefWorks:167}. Recently, we showed that this can implement Bennett clocking in nanomagnetic logic arrays~\cite{RefWorks:154} where a very small voltage of $\sim$16 mV can rotate the magnetization of a shape-anisotropic multiferroic nanomagnet (made of lead zirconium titanate (PZT) and Terfernol-D) by $\sim$90$^{\circ}$ to carry out Bennett clocking. The energy dissipated is only $\sim$60 $kT$ at room temperature and the rotation takes about 80 ns to complete~\cite{fasha10}. The time can be reduced to $\sim$3 ns by increasing the stress to the maximum that PZT might be able to generate in the Terfenol-D layer, but that also increases the energy dissipation to about 48,000 $kT$~\cite{fasha10}. Still, this is considerably smaller than the energy that would have been expended had we clocked the same nanomagnet with spin transfer torque (STT) in the same 3 ns~\cite{fasha10}.

Unfortunately, the voltage-generated-stress (VGS) mode of switching has a major shortcoming. It can rotate an isolated magnet's magnetization by nearly 90$^{\circ}$ (e.g. from the easy to the hard axis of a shape anisotropic magnet) but cannot rotate past 90$^{\circ}$ to achieve a complete flip. Thus, it may be adequate for Bennett clocking which requires rotation by $\sim$90$^{\circ}$, but inadequate for writing a bit in STTRAM, which requires $\sim$180$^{\circ}$ rotation. Therefore, we have devised a mixed-mode approach where both VGS and STT are employed to rotate a magnet's magnetization by $\sim$180$^{\circ}$. We will show that this mixed mode switching (MMS) results in considerable energy saving at any given switching delay compared to using just the STT mode. 
\begin{figure}
\includegraphics[width=2.5in]{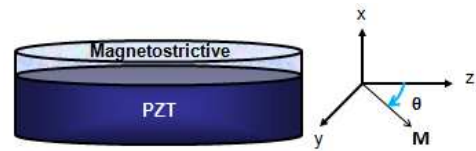}
\caption{\label{fig:multiferroic} An elliptical multiferroic nanomagnet in the y-z plane.}
\end{figure}

In order to study the MMS paradigm, we have solved the appropriate Landau-Lifshitz-Gilbert (LLG) equation analytically. Our nanomagnet is an elliptical multiferroic consisting of a 2 nm/3.5 nm thick layer of Terfenol-D/nickel (magnetostrictive) and a 40 nm thick layer of PZT (piezoelectric) as shown in Fig.~\ref{fig:multiferroic}. Its major and minor axes are 220 nm and 100 nm, respectively, which we tacitly assume precludes formation of multiple domains. This shape anisotropy causes an energy barrier of 32 $kT$ to appear between the two minimum energy states along the major axis (easy axis). We assume that the magnet's plane is the y-z plane and the easy axis is along the z-axis. The angle subtended by the nanomagnet with the +z-axis is denoted as $\theta$ (see Fig. 1). 

Solution of the LLG equation yields the time rate of change of the angle $\theta$. The derivation can be found in the supplementary material, but the final expression is 
\begin{equation}
{{d \theta}\over{dt}} = {{\gamma}\over{\left (1 + \alpha^2 \right )\mu_0 M_s \Omega}} \left\lbrack s \, sin(\xi-\theta) - 2 \alpha B sin\theta cos\theta \right\rbrack,
\label{equation1}
\end{equation}
where $\alpha$ is the Gilbert damping constant of the magnetostrictive layer,
$\Omega$ is the volume of that layer, $\gamma = 2 \mu_B/\hbar$ is the gyromagnetic ratio, $B_0 = (\mu_0/2) \, M_s^2 \Omega \left\lbrack N_{d-yy} - N_{d-zz}\right\rbrack$ (energy barrier due to shape anisotropy), $M_s$ is the saturation magnetization of the magnetic layer, $\mu_0$ is the permeability of free space, $N_{d-yy}$ and $N_{d-zz}$ are the demagnetization factors in the y- and z-directions, respectively, $B_{stress} = (3/2) \lambda_s \sigma \Omega$ (stress-anisotropy energy), $\lambda_s$ is the magnetostrictive coefficient of the polycrystalline magnetic layer, $\sigma$ is the stress, $B = B_0 + B_{stress}$, $s = (\hbar/2e) \eta I$, $I$ is the in-plane spin polarized current inducing STT, $\eta$ is the spin polarization of the current, and $\xi$ is the angle between the spin polarization of the current and the +z-axis. In an actual
STTRAM configuration, where there will be two magnets separated by an insulating layer, the quantity $s$ will be replaced by the quantity $s \lbrace
c(V) - \alpha b(V) \rbrace$, where $b(V)$ and $c(V)$ are dimensionless
voltage-dependent quantities \cite{RefWorks:92}, but for an isolated magnet, they do not arise. 

In order to flip the magnetization from --z to the +z direction, the spins in the spin polarized current must be aligned along the +z-direction so that $\xi = 0$. 

From Equation (\ref{equation1}), we can obtain the time $\tau$ required to rotate the magnetization from an initial orientation $\theta_1$ to a final orientation $\theta_2$ as 
\begin{equation}
\tau =  \int_0^\tau dt = - \int_{\theta_1}^{\theta_2}
{{\left ( 1 + \alpha^2 \right ) \mu_0 M_s \Omega}\over{\gamma
 \left\lbrack s \, sin(\theta) + \alpha B sin (2 \theta) \right\rbrack }} d \theta .
\label{delay}
\end{equation}
We will assume that the initial orientation is aligned close to the --z-axis so that $\theta_{initial} = 180^{\circ} - \epsilon$. If
$\epsilon = 0$ and the magnetization is exactly along the easy axis, then no amount of stress or spin polarized current can budge it since the effective torque exerted on the magnetization by either stress or spin polarized current will be exactly zero. Such locations are called ``stagnation points''. Therefore, we will assume that $\epsilon = 1^{\circ}$. This is not an unreasonable assumption since thermal fluctuations can dislodge the magnetization from the easy axis and make $\epsilon \rightarrow 1^{\circ}$.

We now consider three different scenarios: pure stress mediated switching (VGS), pure spin transfer torque mediated switching (STT) and mixed mode switching (MMS). 

In the pure VGS switching mode, the time taken to rotate the magnetization from the initial orientation $\theta = 180^{\circ} - \epsilon$ to  $\theta = 180^{\circ} - \phi$ ($\phi \leq 90^{\circ}$) is obtained from Equation~\eqref{delay} as
\begin{eqnarray}
t_{VGS} &=& - \left. \frac{\left(1+\alpha^2 \right) \mu_0 M_s \Omega}{2 \gamma  \alpha B}  ln |tan \theta| \right |_{\theta = 180^{\circ} - \epsilon }^{\theta = 180^{\circ} - \phi } \nonumber\\
&=& - \frac{\left(1+\alpha^2 \right) \mu_0 M_s \Omega}{2 \gamma  \alpha B}  ln \left|{{tan \phi}\over{tan \epsilon}}\right| ,
\label{eq:theta_dynamics_stress} 
\end{eqnarray}

The above equation shows that stress will take infinite time to complete the rotation if either $\epsilon = 0$, or $\phi = 90^{\circ}$. Thus, initial alignment along the easy axis or final alignment along the hard axis is forbidden. The physics behind the stagnation points can be understood by looking at the energy profile of the nanomagnet $E (\theta)$ as a function of $\theta$ (see the supplementary material for an expression that relates $E(\theta)$ to $\theta$). Without spin transfer torque, the energy consists of only the shape anisotropy energy and the stress anisotropy energy. The slope of this energy profile $E (\theta)$ at $\theta = 0^{\circ}, 90^{\circ}$ and 180$^{\circ}$ is exactly zero, which means that there is no torque acting on the nanomagnet at these orientations to cause any rotation. This ensures that the pure VGS mode can never rotate the magnetization past 90$^{\circ}$. It can  rotate the magnetization from close to the easy axis to close to the hard axis,  which is adequate for Bennett clocking of nanomagnetic logic~\cite{RefWorks:154}, but not adequate for writing bits in MRAM.

Next, let us consider the pure STT switching mode with no stress present. The time required to rotate the magnetization from an initial orientation $\theta = 180^{\circ} - \vartheta$ to a final orientation $\theta = \vartheta$ is obtained from Equation~\eqref{delay} as (see supplementary material)
\begin{eqnarray}
&& t_{STT}   =  \frac{\left(1+\alpha^2 \right) \mu_0 M_s \Omega}{2 \gamma \, \alpha B_0} \, \frac{m}{1- m^2} \nonumber\\
&& \; \left\lbrack m \, ln \left|\cfrac{1-m cos\,\vartheta}{1+m cos\,\vartheta}\right| - ln \left|\cfrac{1- cos\,\vartheta}{1+ cos\,\vartheta}\right| \right\rbrack ~~ (m \leq 1),
\label{eq:delay_stt}
\end{eqnarray}
where $m = 2 \alpha B_0/s$ and $\vartheta$ is an arbitrary angle. The above expression is invalid for $m > 1$ since then the energy imparted by the spin transfer torque will not be sufficient to overcome the energy barrier due to shape anisotropy and cause rotation of the magnetization. 

The last expression shows that the time to complete the rotation approaches infinity if $\vartheta \rightarrow 0$, indicating that spin transfer torque has a stagnation point at $\theta = 0^{\circ}$ and $180^{\circ}$, i.e. when the final or the initial state of the nanomagnet is along the easy axis. However, there is no stagnation point at the hard axis ($\theta = 90^{\circ}$), unlike in the case of VGS. Hence, thermal fluctuations are needed to nudge the magnetization slightly away from the easy axis so that spin transfer torque can nearly complete the flip. Once again, the origin of the stagnation points can be understood by noting that no spin transfer torque acts on the nanomagnet at these locations. 

If thermal fluctuations deflect the nanomagnet from the easy axis by an angle $\epsilon$, then the time $\Delta t$ needed to complete a near-flip is obtained from Equation (\ref{eq:delay_stt}) by setting $\vartheta = \epsilon$. Note that $\Delta t$ becomes exponentially longer as $\epsilon$ approaches zero, so that one can nearly, but not completely, flip magnetization with STT.

For the MMS mode, we consider the following switching sequence which results in the most energy saving. Stress is applied abruptly to initiate the switching and rotate the magnetization from its initial orientation along $180^{\circ} - \epsilon$  to $\theta = 180^{\circ} - \vartheta$ ($\vartheta < 90^{\circ}$), then a constant spin polarized current is turned on instantaneously. Thereafter, the sign of the stress is reversed abruptly as soon as the hard axis is crossed, and then the spin polarized current is turned off instantaneously when $\theta = \vartheta$. Finally stress is removed when $\theta = \epsilon$. This completes the switching by rotating the nanomagnet from $\theta = 180^{\circ} - \epsilon$ to $\theta = \epsilon$, where $\epsilon = 1^{\circ}$ in our
simulations.
From Equation~\eqref{delay}, the time required to rotate from $180^\circ-\vartheta$ to $\vartheta$ is $t_{MMS} = \tau_{MMS +} + \tau_{MMS -}$, where $\tau_{MMS +}$ and $\tau_{MMS -}$ are given by
\begin{eqnarray}
&& \tau_{MMS\pm} = \pm \frac{\left(1+\alpha^2 \right) \mu_0 M_s \Omega}{ \gamma \, (m^2_\pm-1)} \left\lbrack - 2m_\pm ln|1 \pm m_\pm cos\,\vartheta| \right.\nonumber\\
&& \quad \left. + (m_\pm-1) \, ln |1 \mp cos\,\vartheta| + (m_\pm+1) \, ln |1 \pm cos\,\vartheta| \right\rbrack
\label{eq:delay_stt_stress}
\end{eqnarray}
\noindent
where $m_\pm=2\alpha \left ( B_0 \pm B_{stress} \right )/s$.  The upper sign denotes the switching delay for the rotation from $180^{\circ}-\vartheta$ to $90^{\circ}$ and the lower sign denotes the switching delay for the rotation from $90^{\circ}$ to $\vartheta$. We add to this the time to rotate with stress alone between $\theta=180^\circ-\epsilon$ to $\theta=180^\circ-\vartheta$ and from $\theta=\vartheta$ to $\theta=\epsilon$ which are obtained from Equation~\eqref{eq:theta_dynamics_stress}.

Our next task is to compute the energy dissipations in the different switching modes. Since the pure VGS mode cannot rotate past the hard axis and flip the magnetization, we shall not consider it further and concentrate instead on the pure STT mode and the MMS mode.

In the pure STT mode, we will first pick a switching delay $\Delta t$ and then back out the current $I (\Delta t)$ needed to switch in that time from Equation~\eqref{eq:delay_stt} using appropriate material values for nickel and Terfenol-D given in the supplementary material. The energy dissipated is then computed as $E_{diss}(\Delta t) = I^2(\Delta t) R \Delta t$, where $R$ is the resistance of the magnetostrictive layer. This allows us to plot $E_{diss}$ versus $\Delta t$ for the pure STT mode.

In the MMS mode, the energy dissipated is $E_{diss}(\sigma, \Delta t, \vartheta) = (5/2)CV^2 (\sigma) + I^2(\Delta t, \sigma, \vartheta)R \Delta t$, where $C$ is the capacitance of the piezoelectric layer (estimated as 3.8 fF for the dimensions of the multiferroic nanomagnet and assuming a relative dielectric constant of 1000 for the PZT), $V (\sigma)$ is the voltage applied on the multiferroic to generate a given stress $\sigma$, and $I(\Delta t, \sigma, \vartheta)$ is the spin polarized current needed to rotate the magnetization from $180^{\circ} - \vartheta$ to $\vartheta$ and complete the switching in time $\Delta t$ in the company of that given stress. We pick a stress $\sigma$ and a $\Delta t$, and then back out the current $I(\Delta t, \sigma, \vartheta)$ from Equation~\eqref{eq:delay_stt_stress} that will be needed to complete the switching in time $\Delta t$ while applying STT in the angle interval [180$^{\circ}$ - $\vartheta$, $\vartheta$]. Clearly, $E_{diss}(\sigma, \Delta t, \vartheta)$ depends on $\vartheta$. Therefore, we have to first find $E_{diss}(\sigma, \Delta t, \vartheta)$ for given $\Delta t$ and $\sigma$ as a function of $\vartheta$ (see supplementary material for a plot of $E_{diss}(\sigma, \Delta t, \vartheta)$ versus $\vartheta$). We then choose that value of $\vartheta$ for which the $E_{diss}(\sigma, \Delta t, \vartheta)$ is {\it minimum} since that will yield the most energy efficient switching scheme. For this optimum value of $\vartheta$, we find $E_{diss}(\sigma, \Delta t)$ and repeat this procedure for every $\sigma$, $\Delta t$.

\begin{figure}[htbp]
\centering
\subfigure[]{\label{fig:energy_stress_stt_terfenolD_both}\includegraphics[width=3in]{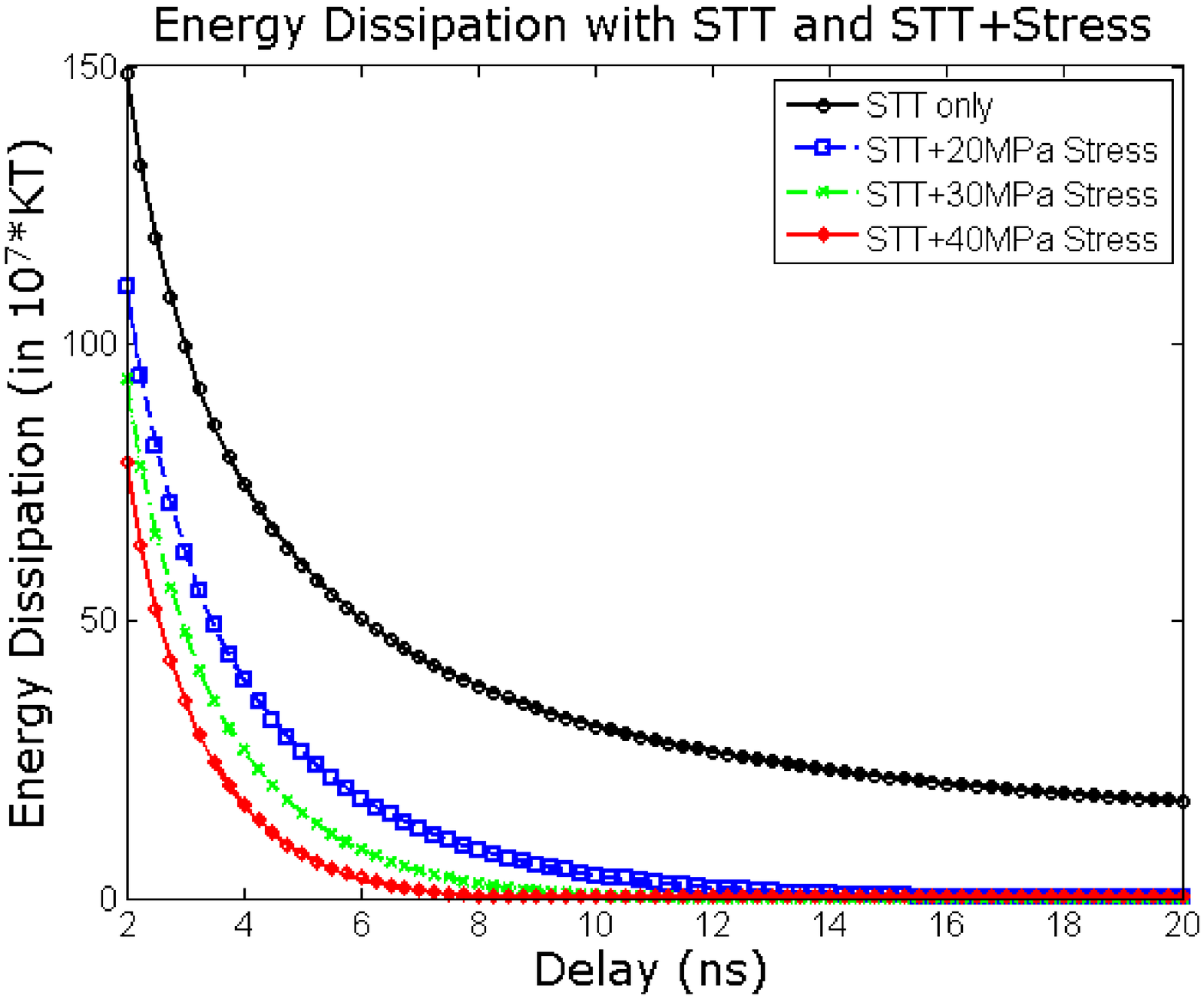}}
\subfigure[]{\label{fig:energy_stress_stt_nickel_both}\includegraphics[width=3in]{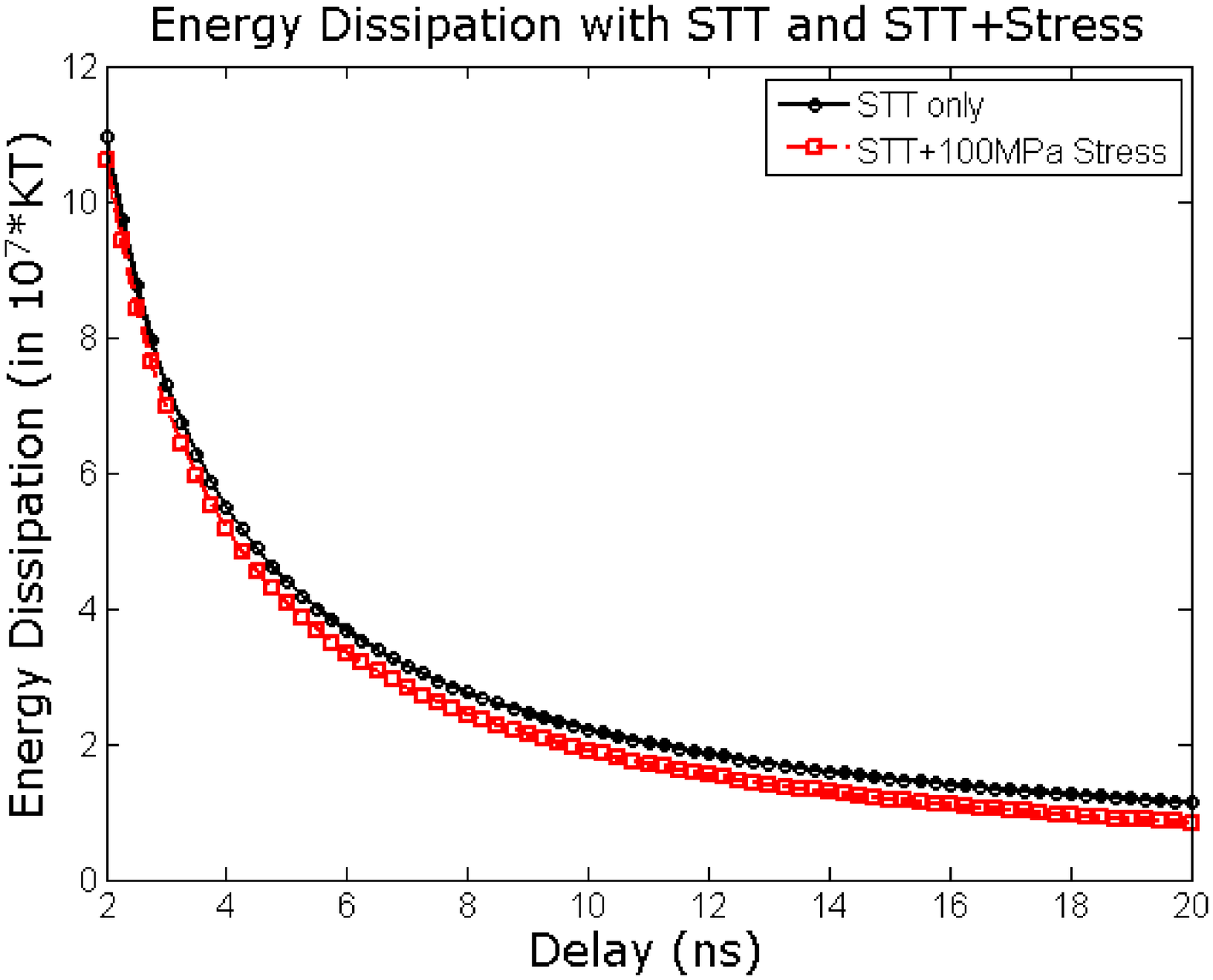}}
\caption{\label{fig:energy_stress_stt} Energy dissipation as a function of switching delay $\Delta t$ for various applied stress $\sigma$ in the MMS mode for (a) Terfenol-D/PZT and (b) nickel/PZT. The plot for the pure STT mode is also shown. }
\end{figure}

In Fig. \ref{fig:energy_stress_stt}, we plot the energy dissipations for the pure STT mode and the MMS mode as a function of delay in both Terfenol-D/PZT and nickel/PZT multiferroics with stress as a parameter. Typical switching delays of magnets contemplated for STTRAM is 2--20 ns~\cite{srcm09}, which is why we plot within this range of delay. In this range, MMS is always superior to STT in the same material, and the energy saving increases with increasing delay since a larger portion of the rotation role can be handed over to stress if we allow for a larger delay. The saving is very pronounced in Terfenol-D/PZT (3 orders of magnitude at long delays) but much less pronounced for nickel (36\% at long delays of 20 ns). However, for delays exceeding 6 ns, the MMS mode in Terfenol-D/PZT is more energy efficient than both the pure STT in nickel and the MMS mode in nickel/PZT.

In conclusion, we have shown MMS mode switching can be considerably more energy-efficient than STT mode. This could potentially lead to a new family of magnetic logic and memory devices based on multiferroics that are switched with a combination of voltage-generated stress and spin polarized current.


\begin{thebibliography}{1}
\providecommand{\bibnamefont}[1]{#1}
\providecommand{\bibfnamefont}[1]{#1}
\providecommand{\url}[1]{\texttt{#1}}
\providecommand{\urlprefix}{URL }
\expandafter\ifx\csname urlstyle\endcsname\relax
  \providecommand{\doi}[1]{doi:\discretionary{}{}{}#1}\else
  \providecommand{\doi}{doi:\discretionary{}{}{}\begingroup
  \urlstyle{rm}\Url}\fi
\providecommand{\eprint}[2][]{\url{#2}}

\bibitem{RefWorks:7}
\bibfnamefont{J.~Z.} \bibnamefont{Sun}, Phys. Rev. B \textbf{62}, 570 (2000).

\bibitem{RefWorks:124}
\bibfnamefont{B.}~\bibnamefont{Behin-Aein}, \emph{et~al.}, IEEE Trans.
  Nanotech. \textbf{8}, 505 (2009).

\bibitem{RefWorks:43}
\bibfnamefont{J.~A.} \bibnamefont{Katine}, \emph{et~al.}, Phys. Rev. Lett.
  \textbf{84}, 3149 (2000).

\bibitem{RefWorks:32}
\bibfnamefont{G.~D.} \bibnamefont{Fuchs}, \emph{et~al.}, Phys. Rev. Lett.
  \textbf{96}, 186603 (2006).

\bibitem{RefWorks:154}
\bibfnamefont{J.}~\bibnamefont{Atulasimha} \emph{et~al.}, Appl. Phys. Lett.
  \textbf{97}, 173105 (2010).

\bibitem{RefWorks:167}
\bibfnamefont{T.}~\bibnamefont{Brintlinger}, \emph{et~al.}, Nano. Lett.
  \textbf{10}, 1219 (2010).

\bibitem{fasha10}
\bibfnamefont{M.~S.} \bibnamefont{Fashami}, \emph{et~al.}, Nanotechnology
  (submitted)  (2010).

\bibitem{RefWorks:92}
\bibfnamefont{D.}~\bibnamefont{Datta}, \emph{et~al.}, Arxiv preprint
  arXiv:0910.2489  (2009).

\bibitem{srcm09}
\bibnamefont{www.src.org/calendar/e003676/FinalReport.pdf}  (2009).

\end{thebibliography}

\end{document}


\maketitle

\section{Magnetization dynamics of a free two-dimensional nanomagnet: Solution of
the Landau-Lifshitz-Gilbert equation}

Consider a two-dimensional isolated nanomagnet of ellipsoidal shape lying in the y-z plane
with its major axis aligned along the z-direction and minor axis along the y-direction
(Fig. 1 of the main paper). The dimension of the major axis is $a$ and that of the minor axis is
$b$, while the thickness is $l$. The volume of the nanomagnet is $\Omega=(\pi/4)a b l$. The 
magnetization of the magnet always lies in the y-z plane since the magnet is two-dimensional.
Let $\theta(t)$ be the angle subtended by the magnetization with the +z-axis
at any instant of time $t$. 

The total energy of the single-domain nanomagnet is the sum of the uniaxial shape anisotropy energy 
and the stress anisotropy energy:
\begin{equation}
E = E_{SHA} + E_{STA},
\end{equation}
\noindent
where $E_{SHA}$ is the uniaxial shape anisotropy energy and $E_{STA}$ is the stress 
anisotropy energy. The former is given by 
\begin{equation}
E_{SHA} = (\mu_0/2) M_s^2 \Omega N_d,
\end{equation}
\noindent
where $M_s$ is the saturation magnetization 
and $N_d$ is the demagnetization factor expressed as 
\begin{equation}
N_d = N_{d-zz} cos^2\theta(t) + N_{d-yy} sin^2\theta(t)
\end{equation}
\noindent
with $N_{d-zz}$ and $N_{d-yy}$ being the components of $N_d$ along the $z$-axis 
and $y$-axis, respectively. 
When $l \ll a,b$, $N_{d-zz}$ and $N_{d-yy}$ are given by~\cite{RefWorks:157}
\begin{subequations}
\begin{eqnarray}
		N_{d-zz} &=& \frac{\pi}{4} \left(\frac{t}{a} \right) 
\left\lbrack 1 - \frac{1}{4}\left(\frac{a-b}{a} \right) - \frac{3}{16}
\left(\frac{a-b}{a} \right)^2 \right\rbrack \\
		N_{d-yy} &=& \frac{\pi}{4} \left(\frac{t}{a} \right) 
\left\lbrack 1 + \frac{5}{4}\left(\frac{a-b}{a} \right) + \frac{21}{16}
\left(\frac{a-b}{a} \right)^2 \right\rbrack.
\end{eqnarray}
\end{subequations}

Note that uniaxial shape anisotropy will favor lining up the magnetization along the 
major axis (z-axis) by minimizing $E_{SHA}$, which is why we will call the major axis the 
``easy axis'' and the minor axis (y-axis) the ``hard axis''. 
We will assume that a force along the z-axis (easy axis) generates stress in the magnet.
In that case, the stress anisotropy energy is given by
\begin{equation}
E_{STA} = - (3/2) \lambda_s \sigma \Omega \, cos^2\theta(t),
\end{equation}
\noindent
where $(3/2) \lambda_s$ is the magnetostriction coefficient of the magnet
and $\sigma$ is the stress generated in it by an external agent. Note that a positive 
$\lambda_s \sigma$ product will favor alignment of the magnetization along the 
major axis (z-axis), while a negative $\lambda_s \sigma$ product will favor alignment 
along the minor axis (y-axis), because that will minimize $E_{STA}$. In our convention, 
a compressive stress is negative and tensile
stress is positive. Therefore, in a material like Terfenol-D that has positive $\lambda_s$,
a compressive stress will favor alignment along the minor axis, and tensile along the major
axis. The situation will be opposite with nickel that has negative $\lambda_s$.
 
At any instant of time, the total energy of the nanomagnet can be expressed as 
\begin{equation}
E(t) = B sin^2\theta(t) + C
\end{equation}
\noindent
where 
\begin{subequations}
\begin{align}
B_0 &= \frac{\mu_0}{2} \, M_s^2 \Omega \left\lbrack N_{d-yy} - N_{d-zz}\right\rbrack \\
B_{stress} &= (3/2) \lambda_s \sigma \Omega \displaybreak[3]\\
B &= B_0 + B_{stress} \\
C &= \frac{\mu_0}{2} M_s^2 \Omega N_{d-zz} - (3/2) \lambda_s \sigma \Omega.
\end{align}
\end{subequations}
Note that $B_0$ is always positive, but $B_{stress}$ can be negative or positive according
to the sign of the $\lambda _s \sigma$ product.

The magnetization \textbf{M}(t) of the magnet has a constant magnitude at any given 
temperature but a variable direction,
so that we can represent it by the vector of unit norm 
$\mathbf{n_m}(t) =\mathbf{M}(t)/|\mathbf{M}| = \mathbf{\hat{e}_r}$ where 
$\mathbf{\hat{e}_r}$ is the unit vector in the radial direction in spherical coordinate 
system represented by ($r$,$\theta$,$\phi$). The other two unit vectors in the 
spherical coordinate system are denoted by $\mathbf{\hat{e}_\theta}$ and 
$\mathbf{\hat{e}_\phi}$ for $\theta$ and $\phi$ rotations, respectively. 

The torque acting on the magnetization within unit volume due to shape and stress anisotropy is
\begin{equation}
T_E (t)= - \mathbf{n_m}(t) \times \nabla E[\theta(t)]= - 2 B sin\theta(t) cos\theta(t)
 \, \mathbf{\hat{e}_\phi} 
\label{stress-torque}
\end{equation}
\noindent
since $\nabla E[\theta(t)] = (\partial E(t)/\partial \theta(t)) \, \mathbf{\hat{e}_\theta}$ 
and $\partial E(t)/\partial \theta(t) = 2 B sin\theta (t) cos\theta(t)$. This immediately shows
that the torque has out-of-plane component.

Passage of a constant spin-polarized current $I$ though the magnet generates an additional 
spin-transfer-torque given 
by~\cite{RefWorks:112,RefWorks:14}
\begin{equation}
T_{STT}(t) = - s \, \mathbf{n_m}(t) \times \left\lbrack a\mathbf{n_m}(t) + b \mathbf{n_s}(t)
 + c (\mathbf{n_m}(t) \times \mathbf{n_s}(t))\right\rbrack
\end{equation}
\noindent	
where $s = (\hbar/2e)\eta I$ is the spin angular momentum deposition per unit time and 
$\eta$ is the degree of spin-polarization in the current $I$. In order to minimize the resistance in the path of the current $I$, it will be applied in-plane ~\cite{RefWorks:8} rather than perpendicular-to-plane, which has been the tradition ~\cite{RefWorks:212}. The coeffcients $b$ and $c$ 
are voltage-dependent dimensionless terms while $a$ is somewhat irrelevant in this 
context since the term involving $a$ vanishes. The unit vector $\mathbf{n_s}$ is in 
the direction of the initial spin polarization of the incident current and lies in the $y$-$z$ plane. For
the sake of simplicity, we will assume that it is time-invariant.
If the current $I$ flows in the $y$-$z$ plane and the spin polarization  
subtends an angle $\xi$ with the positive $z$-axis, then the spin-transfer torque is given 
by
\begin{equation}
T_{STT}(t) = s \, \left\lbrack - b(V) \, sin(\xi-\theta(t)) \, \mathbf{\hat{e}_\phi} + c(V)
 \, sin(\xi-\theta(t)) \, \mathbf{\hat{e}_\theta} \right\rbrack
\label{STT-torque}
\end{equation}
\noindent
where $b(V)$ and $c(V)$ are the voltage-dependent coefficients~\cite{RefWorks:13,RefWorks:15}.

The magnetization dynamics of the single-domain magnet 
under the action of various torques is described by the
 Landau-Lifshitz-Gilbert (LLG) equation as follows.
\begin{equation}
\cfrac{d\mathbf{n_m}(t)}{dt} + \alpha \left(\mathbf{n_m}(t) \times \cfrac{d\mathbf{n_m}(t)}
{dt} \right) = \cfrac{\gamma}{M_V} \left(T_E(t) + T_{STT}(t) \right)
\label{LLG}
\end{equation}
\noindent
where $\alpha$ is the dimensionless phenomenological Gilbert damping constant, $\gamma = 2\mu_B/\hbar$ 
is the gyromagnetic ratio for electrons and is given by $2.21\times 10^5$ (rad.m).(A.s)$^{-1}$, and 
$M_V=\mu_0 M_s \Omega$. In the spherical coordinate system, 
\begin{equation}
\cfrac{d\mathbf{n_m}(t)}{dt} = \theta ' \, \mathbf{\hat{e}_\theta} + sin \theta (t)\, \phi ' 
\,\mathbf{\hat{e}_\phi}.
\end{equation}
\noindent
where $()'$ denotes $d()/dt$. Accordingly,
\begin{equation}
\alpha \left(\mathbf{n_m}(t) \times \cfrac{d\mathbf{n_m}(t)}{dt} \right) = - \alpha sin 
\theta(t) \, \phi'(t) \,\mathbf{\hat{e}_\theta} +  \alpha \theta '(t) \, \mathbf{\hat{e}_\phi}
\end{equation}
\noindent
and 
\begin{equation}
\cfrac{d\mathbf{n_m}(t)}{dt} + \alpha \left(\mathbf{n_m}(t) \times \cfrac{d\mathbf{n_m}(t)}
{dt} \right) = (\theta '(t) - \alpha sin \theta (t)\, \phi'(t)) \, \mathbf{\hat{e}_\theta} 
+ 
(sin \theta(t) \, \phi ' (t) + \alpha \theta '(t)) \,\mathbf{\hat{e}_\phi}.
\end{equation}
\noindent
Equating the $\hat{e}_\theta$ and $\hat{e}_\phi$ components in both sides of Equation
(\ref{LLG}), we get
\begin{eqnarray}
\theta ' (t) - \alpha sin \theta(t) \, \phi'(t)  &=&  \frac{\gamma}{M_V} s \, c(V) \, 
sin(\xi-\theta(t)) \nonumber\\
sin \theta (t) \, \phi '(t) + \alpha \theta '(t) &=& \frac{\gamma}{M_V} \left( - 
2 B sin\theta(t) cos\theta(t) - s \, b(V) \, sin(\xi-\theta(t)) \right). \nonumber\\
\end{eqnarray}
\noindent
Eliminating $\phi'(t)$ from the last two equations, we get the following equation
involving only $\theta(t)$:
\begin{eqnarray}
\left(1+\alpha^2 \right) \theta'(t) &=& \frac{\gamma}{M_V} \left\lbrack s \, c(V) \, 
sin(\xi-\theta(t)) + \alpha \left( - 2B sin\theta (t)cos\theta (t)- s \, b(V) \, 
sin(\xi-\theta(t)) 
\right) \right\rbrack \nonumber \\
&=& \frac{\gamma}{M_V} \left\lbrack s \left\lbrace c(V) - \alpha \, b(V) \right\rbrace \, 
sin(\xi-\theta(t)) - 2 \alpha B sin\theta(t) cos\theta(t) \right\rbrack.
\label{eq:theta_dynamics}
\end{eqnarray}

In the rest of the derivation, we will assume that the initial magnetization of the 
nanomagnet is close to the negative z-axis, so that the initial value of $\theta$ is close
$180^o$. Our intent is to switch it, i.e. flip it up, so that the final value of
$\theta$ will be close to $0^o$. Accordingly, the spin polarization of the current needs to be 
along the positive-$z$ direction, which means that $\xi=0^o$. Note that we said that the 
initial and the final orientations of the magnetization are ``close to'' 
but not ``exactly along'' the easy axis, since the torques $T_E$ and $T_{STT}$ both tend to
zero when $\theta(t)$ tends to either 0 or $\pi$ as can be easily verified 
from Equations (\ref{stress-torque}) and (\ref{STT-torque}). Since the torques become vanishingly small near the 
easy axis, neither the initial nor the final orientations can be {\it exactly} along the easy axis. 
In fact,
if the initial orientation is exactly along the easy axis, then no torque acts on the magnetization
vector and nothing
can budge it from this location.
The locations where the torque vanishes are called ``stagnation points''. Clearly, the easy axis
is a stagnation point for both VGS and STT mode switching.
Note that the torque 
$T_E$ is zero when $\theta(t) = \pi/2$ as well, so that the hard axis is also a stagnation point for 
the VGS mode, but not the STT mode. We will revisit all this later.

\subsection{Switching delay for stress-mediated rotation or for the pure VGS mode}

In the absence of spin-transfer torque, Equation~\eqref{eq:theta_dynamics} becomes
\begin{equation}
\left(1+\alpha^2 \right) \theta'(t) = - \frac{\gamma}{\mu_0 M_s \Omega} \; 2 \alpha B 
sin\theta(t) cos\theta(t).
\end{equation}
which yields
\begin{equation}
\int dt = - \int {{\left(1+\alpha^2 \right) \mu_0 M_s \Omega }\over{2 \gamma \alpha B 
sin\theta(t) cos\theta(t)}} \, d \theta .
\end{equation}

Integrating, we get that the time $t$ taken for the magnetization angle $\theta(t)$ to change from
$\theta_1$ to $\theta_2$ under the action of stress is
\begin{equation}
t_{VGS} = -\frac{\left(1+\alpha^2 \right) \mu_0 M_s \Omega}{2\gamma \, \alpha B} \, 
ln \left|{{tan \theta_2}\over{tan \theta_1}}\right|.
\label{VGS-time}
\end{equation}

The above expression shows immediately that we can approach, but never actually reach, 
the hard axis with
just stress acting on the magnet (i.e. in the pure VGS mode) since $t_{VGS} \rightarrow
\infty$ as $\theta_2 \rightarrow \pi/2$. In other words, it will take infinite time to reach 
the hard axis starting from {\it any} orientation. It should be also
apparent that it will take infinite time to get out of alignment with the easy axis if the initial
orientation was along the easy axis since $t_{VGS} \rightarrow \infty$ as $\theta_1 
\rightarrow \pi$. Therefore, clearly, both easy and hard axes are stagnation points. The switching time 
tends to infinity in these cases since the torque vanishes at the stagnation points, as we have 
seen already.
As a result, the VGS mode may be good for Bennett clocking \cite{RefWorks:154} which merely
requires rotating magnetization from near the easy axis to near the hard axis, but it is 
useless for flipping magnetization needed in spin valves and magnetic tunnel junctions,
since it is impossible to rotate past the hard axis and complete a 
near-180$^{\circ}$ rotation$\footnote{Bennett clocking in nanomagnetic 
logic arrays is also greatly facilitated by the dipole interaction of near neighbors which
comes to the aid of stress. Discussion of this issue is however outside
the scope of this paper.}$.

\subsection{Switching delay for spin-transfer torque (STT)}

In absence of stress, Equation~\eqref{eq:theta_dynamics} becomes (assuming $\xi = 0$)
\begin{equation}
\left(1+\alpha^2 \right) \theta' (t)= -\frac{\gamma}{M_V} \left\lbrack s' \, sin\theta(t)
 + 2 \alpha B_0 sin\theta (t)cos\theta (t)\right\rbrack
\label{eq:theta_dynamics_stt}
\end{equation}
\noindent
where $s' = s \left \lbrace c(V) - \alpha b (V) \right \rbrace$ and $B=B_0$ since $B_{stress}=0$. The last equation can be integrated using the 
identity
\begin{eqnarray}
&& \int{\frac{d\theta}{sin\theta (t)+ m sin\theta (t)cos\theta}} = \frac{1}{2(m^2-1)} \nonumber\\
&& \quad \times
\left\lbrack (m-1) \, ln (1-cos\theta(t)) + (m+1) \, ln (1+ cos\theta(t)) - 2  m \, 
ln |1+m\, cos\theta(t)|\right\rbrack.
\label{eq:integration}
\end{eqnarray}
which yields the switching delay associated with rotating the magnetization from $\theta
=\pi-\vartheta$ to $\theta = \vartheta$ as
\begin{equation}
t_{STT} = - \frac{\left(1+\alpha^2 \right) \mu_0 M_s \Omega}{2 \gamma \, \alpha B_0} \, 
\frac{m}{m^2-1} \left\lbrack m \, ln \left|\cfrac{1-m cos\,\vartheta}{1+m cos\,\vartheta}
\right| - ln \left|\cfrac{1- cos\,\vartheta}{1+ cos\,\vartheta}\right| \right\rbrack
\label{eq:delay_stt}
\end{equation}
\noindent
where $m=2\alpha B_0/s'$. The above equation is valid only for $m \leq 1$. If $m > 1$, then the spin transfer torque 
energy is insufficient to overcome the energy barrier due to shape anisotropy
and cause rotation past the hard axis.

\begin{figure}
\centering
\includegraphics[width=6in]{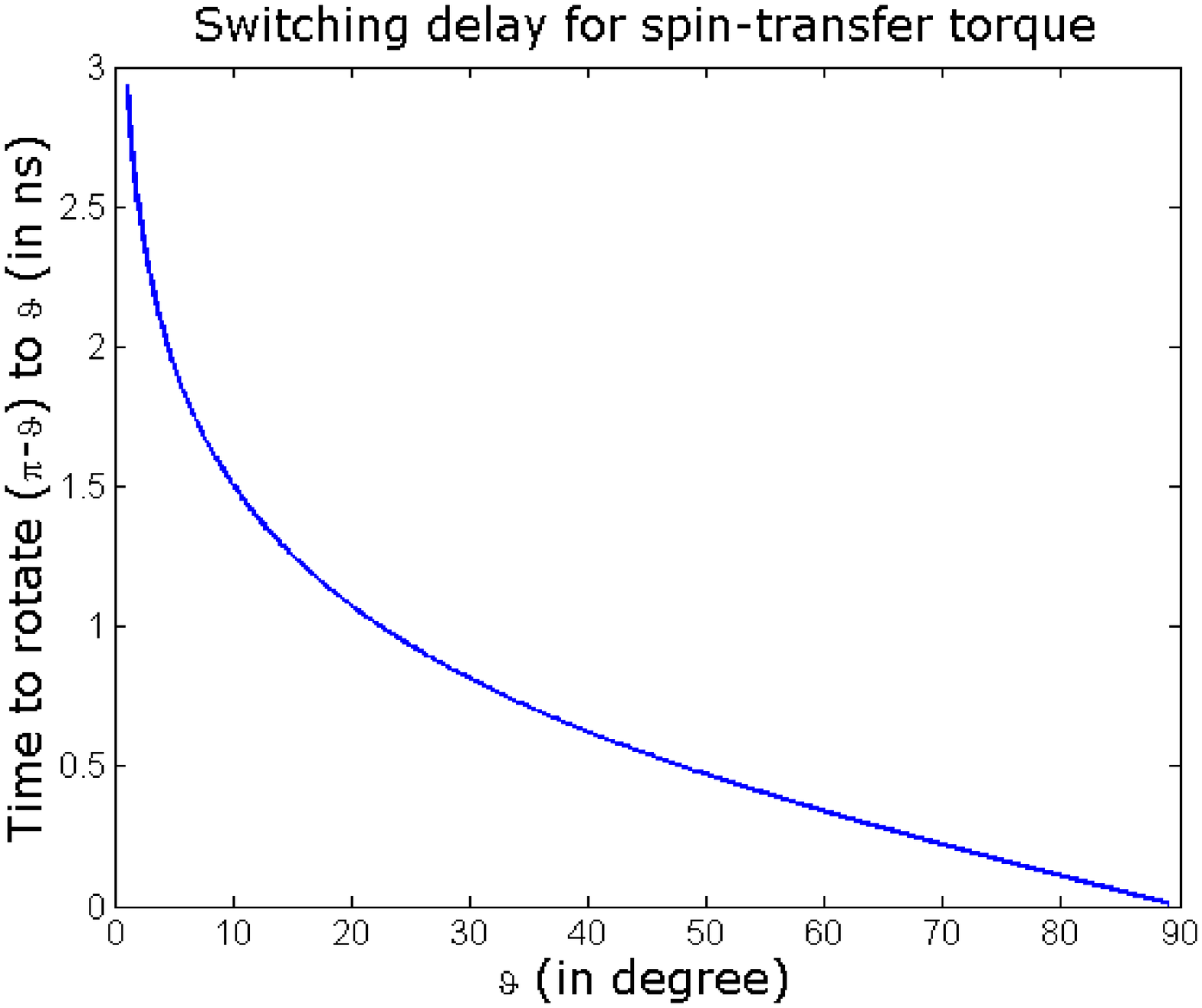}
\caption{\label{fig:delay_stt} Plot of the time taken by STT to rotate magnetization through an
angle $\vartheta$ versus $\vartheta$. The material is Terfenol-D. The initial orientation is assumed to be along the easy axis.
Note that slope is high near $\vartheta = 0$, indicating that it takes a very long time 
to rotate even by a small angle away from the easy axis since that location is a stagnation
point.}
\end{figure}

Figure~\ref{fig:delay_stt} plots the time it takes to rotate the magnetization from 
$\theta = \pi-\vartheta$ to $\theta = \vartheta$ as a function of the angle $\vartheta$.
It is obvious from Equation (\ref{eq:delay_stt}) that the delay should increase rapidly as the initial
oientation of the magnetization approaches the \emph{easy} axis 
since $t_{STT}$ diverges when $\vartheta$ 
approaches 0 or $\pi$ (because of the last logarithmic term in Equation (\ref{eq:delay_stt}). 
This feature is apparent in the
plot as well. Once again this happens because the spin transfer torque vanishes 
when the magnetization is exactly along the easy axis.
Therefore, the easy axis presents a stagnation point and thermal fluctuations
will be needed to get out of it.
However, there is no stagnation point around the hard axis (unlike in the case of VGS
mode switching) which can be verified from Equation~\eqref{eq:theta_dynamics_stt} as well.
This is what allows STT mode switching to rotate the magntization past the hard axis.

The reader can easily ascertain that the time for flipping from negative to positive z-axis
is the same as that for flipping from the positive to the negative z-axis. In the latter case, 
we merely have to make 
$\xi = \pi$ instead of 0, and interchange $\pi - \vartheta$ and $\vartheta$.

\subsection{Switching delay for the mixed mode}

\begin{figure}
\centering
\includegraphics[width=6in]{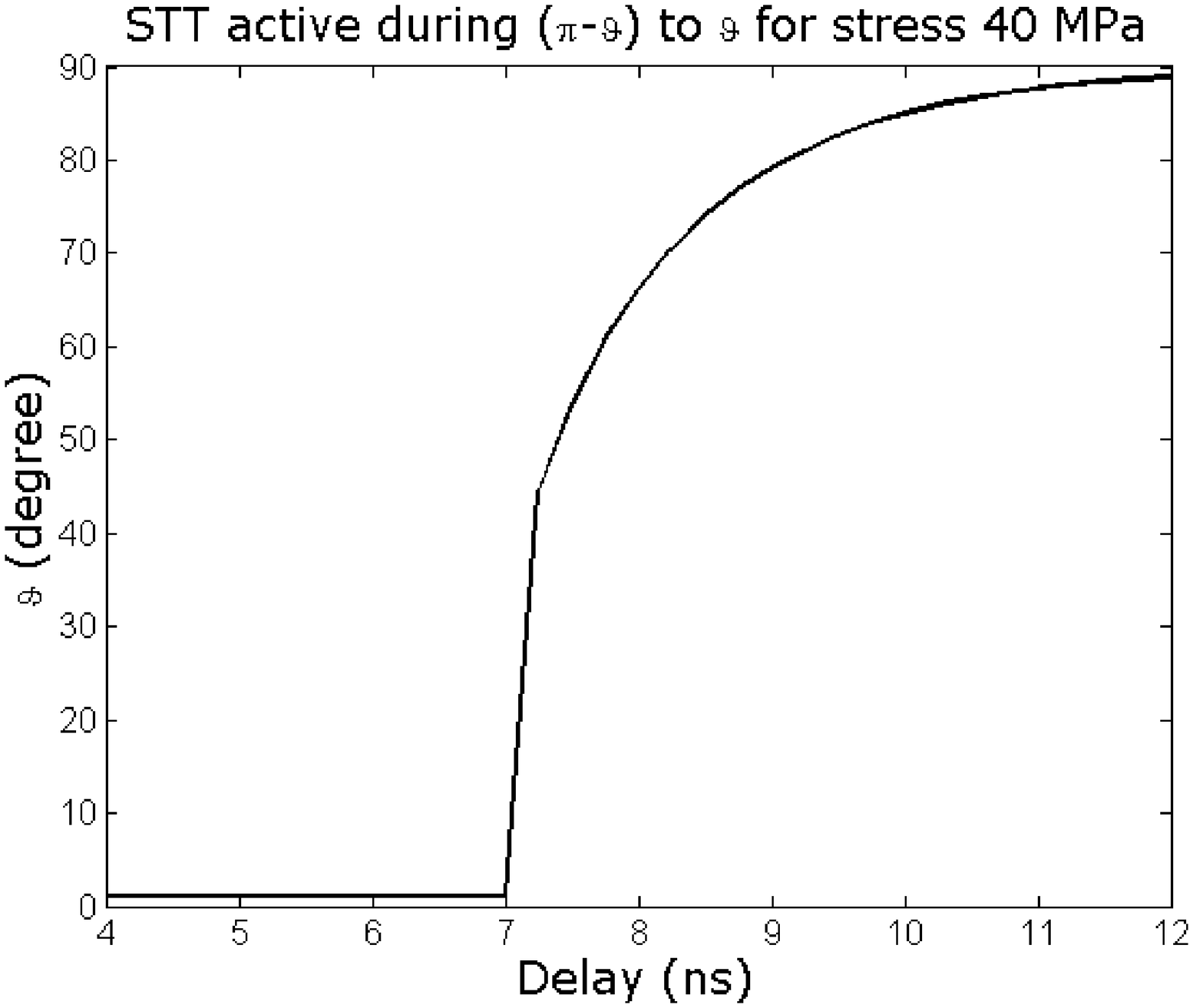}
\caption{\label{fig:vartheta_vs_delay} Plot of the angle $\vartheta$ as a function of the total switching delay $\Delta t$ in the MMS mode for Terfenol-D for a stress of 40 MPa. The angle $\vartheta$ asymptotically approaches 90$^{\circ}$ for very long delays, but can never actually reach 90$^{\circ}$, because then no STT mode would have been required to accomplish the switching. That is impossible since the VGS mode by itself cannot rotate past the hard axis and complete switching.}
\end{figure}

The following considerations come into play in choosing the mixed mode of switching.
The VGS mode cannot rotate past the hard axis and therefore will need assistance from the STT mode
to accomplish this feat. Therefore, the spin polarized current should be turned on when the magnetization approaches the hard axis and turned off after crossing the hard axis, i.e. the current $I$ should be on between $\theta = \pi - \vartheta$ and 
$\theta = \vartheta$. But how should one determine the optimum $\vartheta$? 
Clearly there is an optimum value of $\vartheta$ for a given switching delay since if $\vartheta$ is too 
large and close to $90^\circ$, then we would have burnt up a lot of the
time with stress which is slow and now we will need to pass a very high spin polarized current $I$ to complete 
the switching in the little time remaining. This will be counter-productive since the large current 
will dissipate too much energy. On the other hand, if $\vartheta$ is too small, then we are not 
being sufficiently frugal in the use of energy-inefficient spin transfer torque and once again expending too much energy 
because 
we are running the current $I$ for too long. This too is counter productive. Hence, there is an optimum value of 
$\vartheta$ for a given delay
that must be found from exhaustive search. If the delay is too short, then $\vartheta$ may actually turn out of be zero, or rather $\epsilon$, i.e. the spin polarized current must be on for the entire duration to achieve the switching within the stipulated delay. A plot of $\vartheta$ versus the switching delay $\Delta t$ is given in Fig.~\ref{fig:vartheta_vs_delay}. Note that for short delays, $\vartheta$ is indeed $\epsilon$ ( = 1$^{\circ}$), which means that  the spin polarized current is on for the entire switching duration, and then $\vartheta$ increases sharply with increasing delay. Clearly, the total energy dissipated in the switching process will depend on $\vartheta$ since that determines the duration for passing a spin polarized current. For a given total delay $\Delta t$ and a given stress $\sigma$, one must compute the total
energy dissipated in both the spin polarized current and the stress as a function of $\vartheta$ and choose that value of $\vartheta$
for which the total energy dissipated is minimum.

The stress too must be handled judiciously in the mixed mode. One should remember that 
a positive $\lambda_s \sigma$ product
favors aligning the magnetization along the easy axis while a negative product favors alignment along the 
hard axis in the configuration we choose. Since we are starting off from a location close to the easy axis {\it towards} the hard
axis, we should start off with a negative product (compressive stress for a material
with positive magnetostriction and tensile stress for a material with negative magnetostriction)
and then reverse the stress to make the product positive once the rotation goes past the 
hard axis, because then we will be aiming for the easy axis. This will be the optimum switching strategy.

Switching via the MMS mode is therefore accomplished in four distinct phases: in phase 1, stress of appropriate sign (compressive or tensile) is applied
to rotate from 
$\theta = \pi - \epsilon$ to $\theta = \pi - \vartheta$, then in phase 2, STT is added to stress
to rotate from $\theta = \pi - \vartheta$ to $\theta = \pi/2$, then in phase 3, the sign of the stress is
reversed while still maintaining STT to rotate from $\theta = \pi/2$ to $\theta = \vartheta$, and finally 
in phase 4, 
STT is turned off and stress alone rotates the magnetization from  
$\theta = \vartheta$ to $\theta = \epsilon$.
This nearly completes 
a flip. 

The time taken to complete the first phase of the rotation from $\theta = \pi - \epsilon$ to 
$\theta = \pi - \vartheta$
is found from Equation (\ref{VGS-time})
\begin{equation}
t_{phase1} = -\frac{\left(1+\alpha^2 \right) \mu_0 M_s \Omega}{2\gamma \, \alpha 
\left [B_0 + B_{stress} \right ]} \, 
ln \left|{{tan \vartheta}\over{tan \epsilon}}\right|.
\end{equation}
Note that in the above equation, $\vartheta > \epsilon$. Therefore, to ensure that 
$t_{phase1}$ is positive and hence meaningful, we have to ensure that 
$B_{stress}$ is negative (i.e. the $\lambda_s \sigma$ product is negative) and that
$\left |B_{stress} \right | > B_0$. The former condition determines the sign of stress (compressive or tensile) and the latter condition ensures that 
the stress anisotropy energy can overcome the shape anistropy energy and initiate rotation.

The time taken to complete the final phase of the rotation from $\theta = \vartheta$
to $\theta = \epsilon$ is also found from
Equation (\ref{VGS-time}):
\begin{equation}
t_{phase4} = -\frac{\left(1+\alpha^2 \right) \mu_0 M_s \Omega}{2\gamma \, \alpha 
\left [B_0 + B_{stress} \right ]} \, 
ln \left|{{tan \epsilon}\over{tan \vartheta}}\right|.
\end{equation}

Clearly $t_{phase4}$ will become negative (unphysical) if $B_{stress}$ is negative 
and exceeds $B_0$ in magnitude. To prevent this, the sign of the stress must be opposite to that
during phase 1  so that $B_{stress}$ positive. That will keep $t_{phase4}$ positive. 
Interestingly, a negative $B_{stress}$ is 
still physical (and keeps $t_{phase4}$ positive) as long as its magnitude is less than that 
of $B_0$. In that case, the shape
anisotropy will ensure that the magnet comes to rest along the easy axis in spite of  
opposition from the stress anisotropy energy because shape anisotropy energy 
is stronger. Nonetheless, the opposition from the stress anisotropy energy will slow the 
rotation. In order to speed up the rotation, it is therefore necessary to reverse the stress during 
 phase 4 so that shape anisotropy and stress anisotropy energy work in concert instead 
of opposing each other.

During the second and third phases of the rotation, both VGS and STT modes coexist. The only difference
between these two phases is that during the second phase, $B_{stress} = - \left | B_{stress} \right |$,
while during the third phase, $B_{stress} = \left | B_{stress} \right |$.

In the presence of both stress and spin-transfer torque, Equation~\eqref{eq:theta_dynamics} becomes 
(assuming $\xi=0$)
\begin{equation}
\left(1+\alpha^2 \right) \theta' = -\frac{\gamma}{\mu_0 M_s \Omega} \left\lbrack s' \, 
sin\theta + 2 \alpha B sin\theta cos\theta \right\rbrack
\label{eq:theta_dynamics_stt_stress}
\end{equation}
or
\begin{equation}
\int dt = - \int {{\left (1 + \alpha^2 \right )\mu_0 M_s \Omega}\over{\gamma\left\lbrack s' \, sin\theta + 2 \alpha B sin\theta cos\theta \right\rbrack}} = - \int {{\left (1 + \alpha^2 \right )\mu_0 M_s \Omega}\over{\gamma\left\lbrack s' \, sin\theta + \alpha B sin(2\theta) \right\rbrack}} \, d \theta
\end{equation}
\noindent
where $s'=s  \left\lbrace c(V) - \alpha \, b(V) \right\rbrace$. This is Equation (2) in the main paper.

The above equation immediately shows that stress (contained in the second term in the denominator) is most effective when $\theta=45^\circ$, while the spin transfer torque (contained in the first term in the denominator) is most effective when $\theta=90^\circ$. The stress term actually vanishes when $\theta=90^\circ$ consistent with the fact that stress is ineffective around the hard axis.

We will carry out the integration using the identity in Equation~\eqref{eq:integration} and arrive at the following expression for the 
switching delay: 
\begin{eqnarray}
t_{phase2} & = &  \tau_{MMS +} \nonumber \\
t_{phase3} & = & \tau_{MMS-} .
\end{eqnarray}
\noindent
with
\begin{eqnarray}
\tau_{MMS\pm} &=& \pm \frac{\left(1+\alpha^2 \right) \mu_0 M_s \Omega}{2 \gamma \, 
\alpha (B_0\mp \left |B_{stress} \right |)} \, \frac{m_\pm}{m^2_\pm-1} \nonumber\\
&& \left\lbrack (m_\pm-1) \, ln |1 \mp cos\,\vartheta| + (m_\pm+1) \, 
ln |1 \pm cos\,\vartheta|  - 2m_\pm ln|1 \pm m_\pm cos\,\vartheta| \right\rbrack
\label{eq:delay_stt_stress}
\end{eqnarray}
\noindent
and $m_\pm=2\alpha (B_0 \mp \left |B_{stress} \right |)/s'$. 

\section{Material parameters}

The material parameters that are used in the simulation are given in the Tables~\ref{tab:nickel} and~\ref{tab:terfelonD}~\cite{RefWorks:179,RefWorks:176,RefWorks:178,RefWorks:213, RefWorks:172}.

\begin{table}[htbp]
\centering
\caption{\label{tab:nickel}Material parameters for nickel}
\begin{tabular}{c||c}
\hline \hline
Young's modulus (Y) & 2$\times$10$^{11}$ Pa \\
\hline
Magnetostrictive coefficient ($(3/2)\lambda_s$) & -3$\times$10$^{-5}$ \\
\hline
Saturation magnetization ($M_s$) & 4.84$\times$10$^5$ A/m \\
\hline
Gilbert damping ($\alpha$) & 0.045 \\
\hline
Resistivity ($\rho$) & 7.8$\times$10$^{-8}$ ohm-m \\
\hline\hline
\end{tabular}
\end{table}

\begin{table}[htbp]
\centering
\caption{\label{tab:terfelonD}Material parameters for Terfenol-D}
\begin{tabular}{c||c}
\hline \hline
Young's modulus (Y) & 8$\times$10$^{10}$ Pa \\
\hline
Magnetostrictive coefficient ($(3/2)\lambda_s$) & +90$\times$10$^{-5}$ \\
\hline
Saturation magnetization ($M_s$) & 8$\times$10$^5$ A/m \\
\hline
Gilbert damping ($\alpha$) & 0.1 \\
\hline
Resistivity ($\rho$) & 63.1$\times$10$^{-8}$ ohm-m\\
\hline \hline
\end{tabular}
\end{table}


\section{Procedure for determining the voltage required to generate a given stress in a magnetostrictive 
material}

In order to generate a stress $\sigma$  in a magnetostrictive layer, the strain in that material must 
be $\varepsilon = \sigma/Y$, where $Y$ is the Young's modulus of the material. We will assume that a voltage
applied to the PZT layer strains it and since the PZT layer is much thicker than the magnetostrictive layer,
all the strain generated in the PZT layer is transferred completely to the magnetostrictive layer. Therefore,
the strain in the PZT layer must also be $\varepsilon$. The voltage needed to generate this strain is calculated from
the piezoelectric coefficient $d_{31}$ of PZT.

We assume that the maximum strain that can be produced in the PZT layer and transferred to the magnetostrictive
layer is 500 ppm. This gives the maximum stress in Terfenol-D as 40 MPa and in nickel as 100 MPa.

\section{Simulation results}
\begin{figure}
\centering
\includegraphics[width=6in]{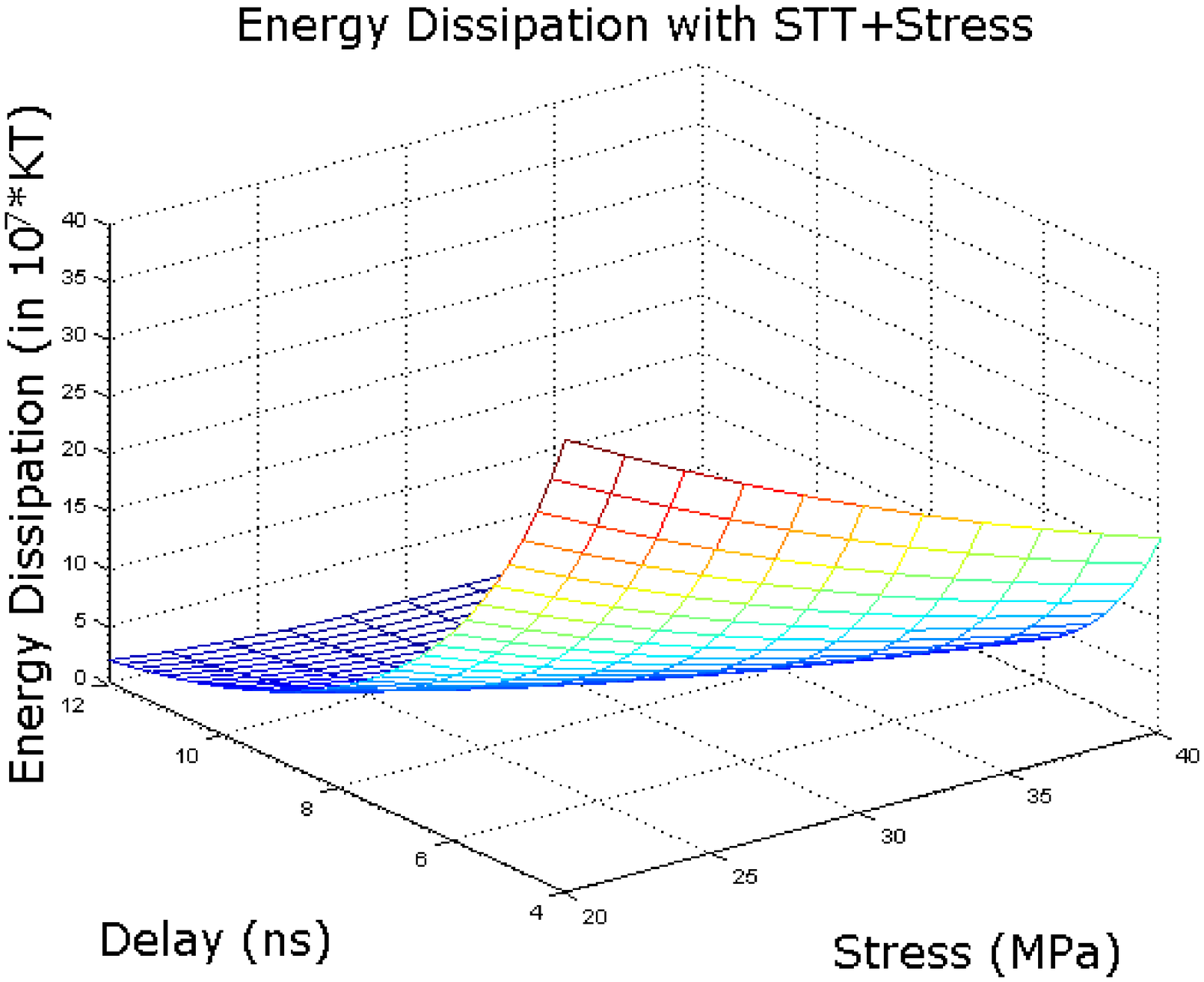}
\caption{\label{fig:energy_stress_stt_mesh} Energy dissipation at different delays 
and at different values of applied stress for the MMS mode, assuming that the material is
Terfenol-D. These quantities are calculated 
for the optimum angle $\vartheta$ to get the minimum energy dissipation.}
\end{figure}

\begin{figure}
\centering
\includegraphics[width=6in]{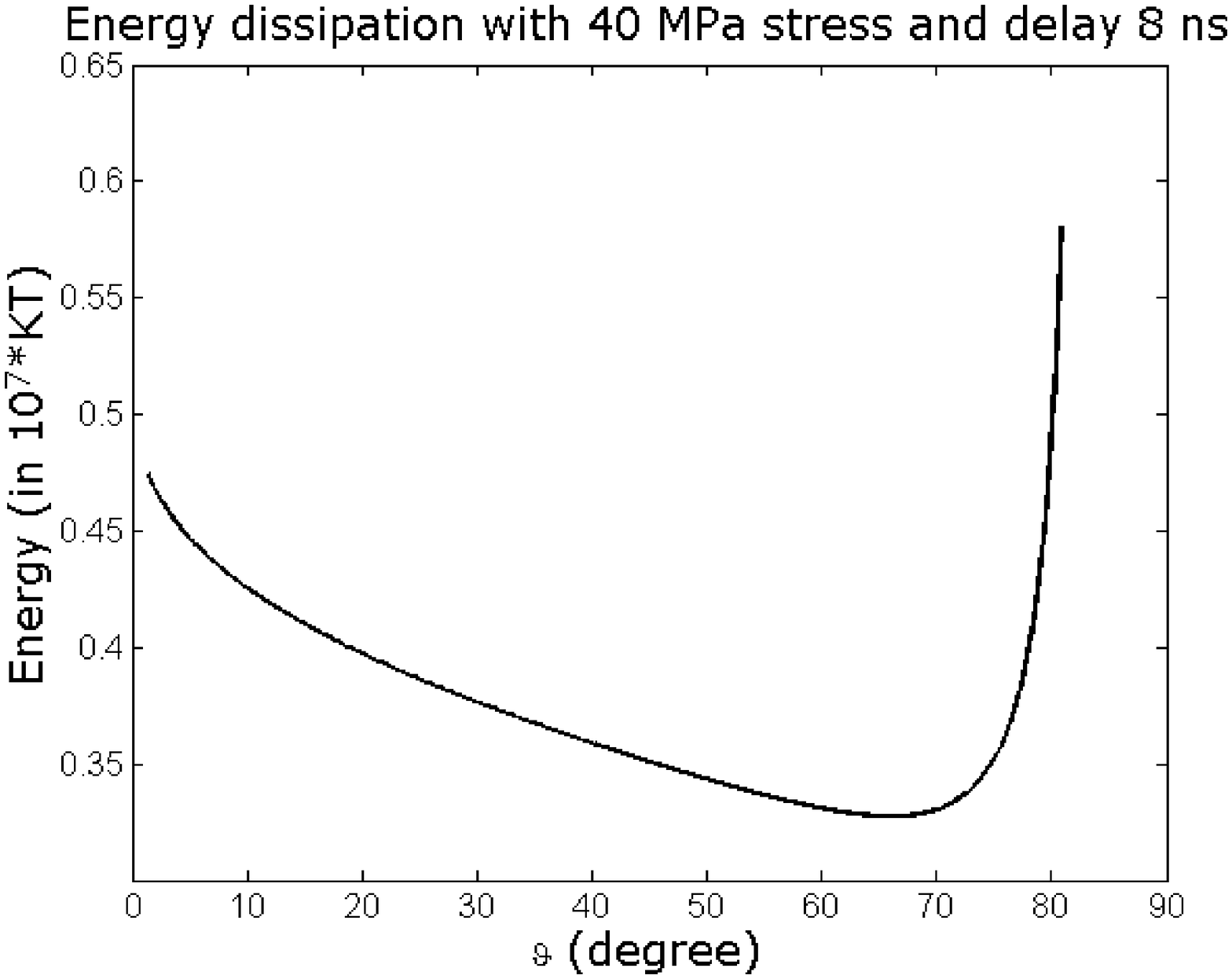}
\caption{\label{fig:optimization} Energy dissipation for a fixed delay of 8 ns 
and for a fixed stress of 40 MPa in Terfenol-D as a function of the angle $\vartheta$ in
the MMS switching mode.}
\end{figure}

\begin{figure}
\centering
\includegraphics[width=6in]{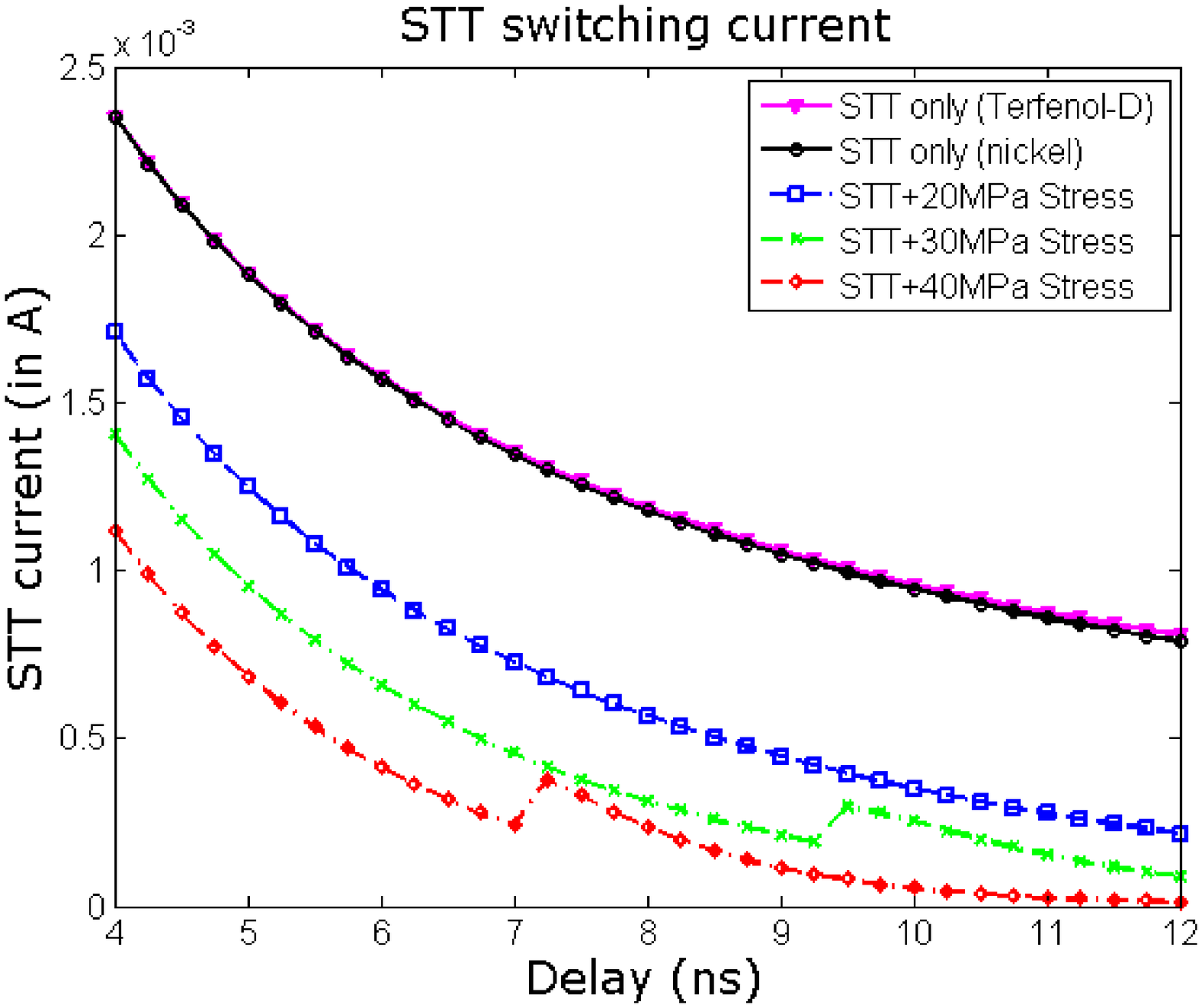}
\caption{\label{fig:current_stress_stt} STT current required to switch using the MMS mode as a function of switching 
delay for different applied stress in the case of Terfenol-D. The current required to switch using the pure STT mode is also shown for both Terfenol-D and nickel as a function of delay. Note that the latter current is almost the same for both Terfenol-D and nickel at any delay within the range shown. At any given delay, the current needed to switch using the MMS mode decreases with increasing stress because stress carries out part of rotation and relegates less and less of the rotation task to STT. At high enough stresses there are sudden jumps in the current when the delay exceeds a certain value. At these delay values, the angle $\vartheta$ is no longer stuck at $\epsilon$ and jumps to a large value (see Fig. \ref{fig:vartheta_vs_delay}) so that STT need not be kept on during the 
entire duration of rotation. It becomes more economical to turn on the STT current only for part of the rotation (from $\theta = 180^{\circ} - \vartheta$ to
$\theta = \vartheta$), but that also requires a larger STT current, which is why we observe the jump.}
\end{figure}

\begin{figure}
\centering
\includegraphics[width=6in]{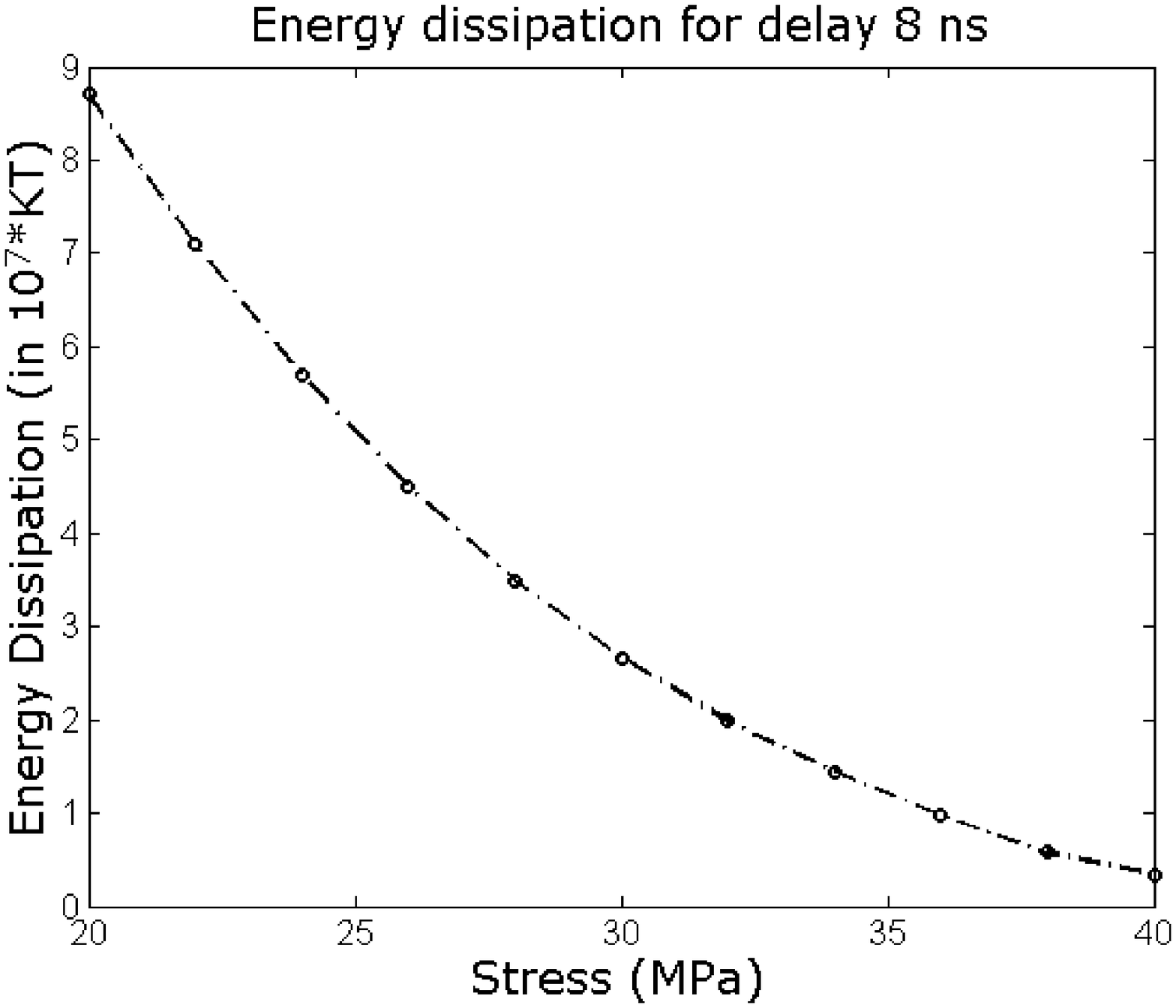}
\caption{\label{fig:energy_total_stress} Total energy dissipated (in stress and spin polarized current) to switch the magnetization of a Terfenol-D nanomagnet using the MMS mode in a fixed delay 
of 8 ns as a function of applied stress. As we increase stress, we require progressively 
less spin transfer current to complete the switching in the stipulated 8 ns. This results in overall energy saving since spin polarized current is far more dissipative than stress. Energy is saved by maximizing the proportion of stress and minimizing the proportion of spin transfer torque current when the switching delay is fixed at 8 ns.
}
\end{figure}

\begin{figure}
\centering
\includegraphics[width=6in]{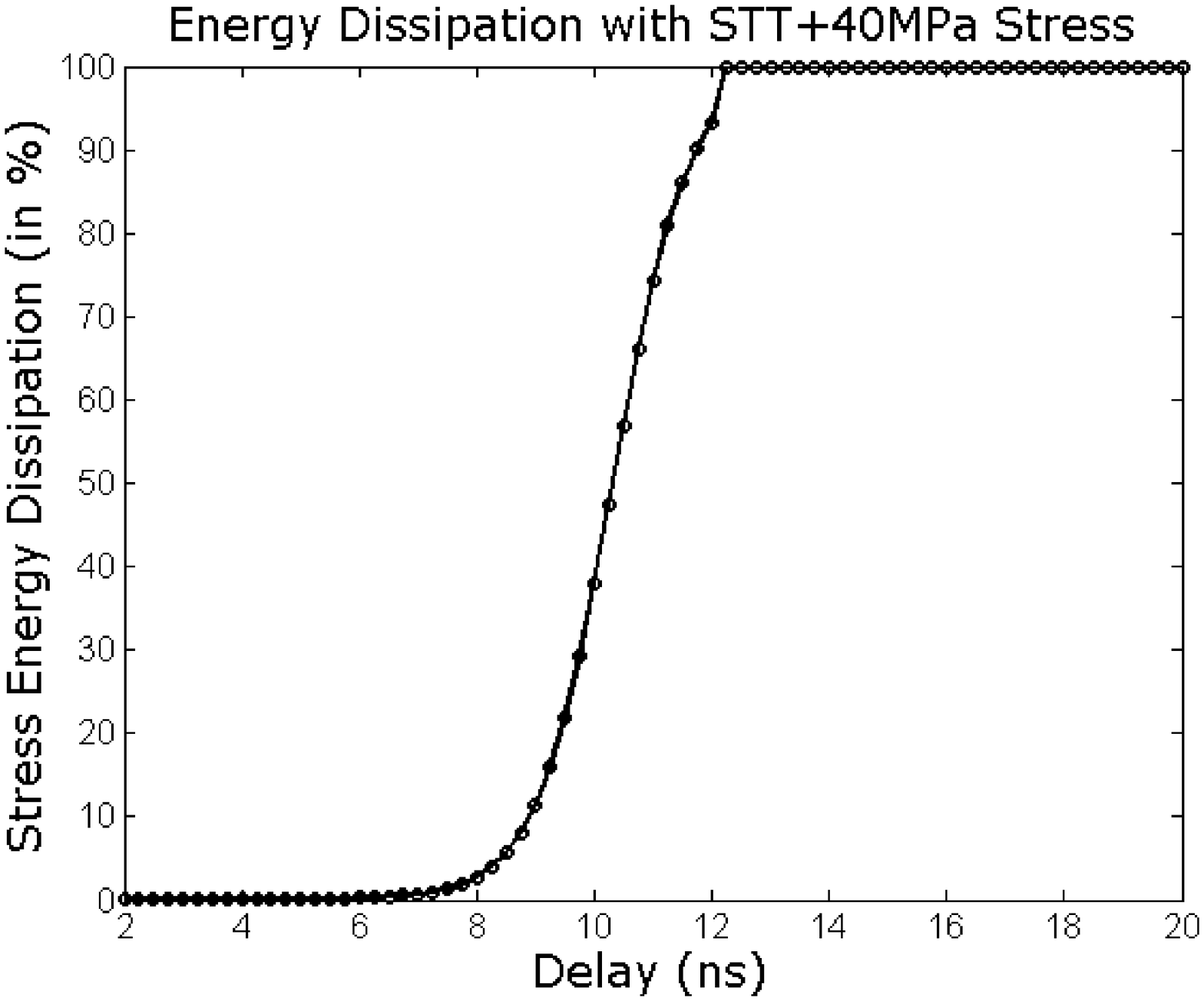}
\caption{\label{fig:energy_stress_percent} Percentage of the total energy dissipation that is due to stress in the MMS mode as a function of switching delay. With increasing delay, we can apportion more of the switching role to stress and less to spin transfer torque so that the percentage of energy consumed by stress increases with delay. The total energy dissipated of course decreases with increasing delay since spin transfer torque current is much more dissipative than stress.}
\end{figure}

\begin{figure}
\centering
\includegraphics[width=6in]{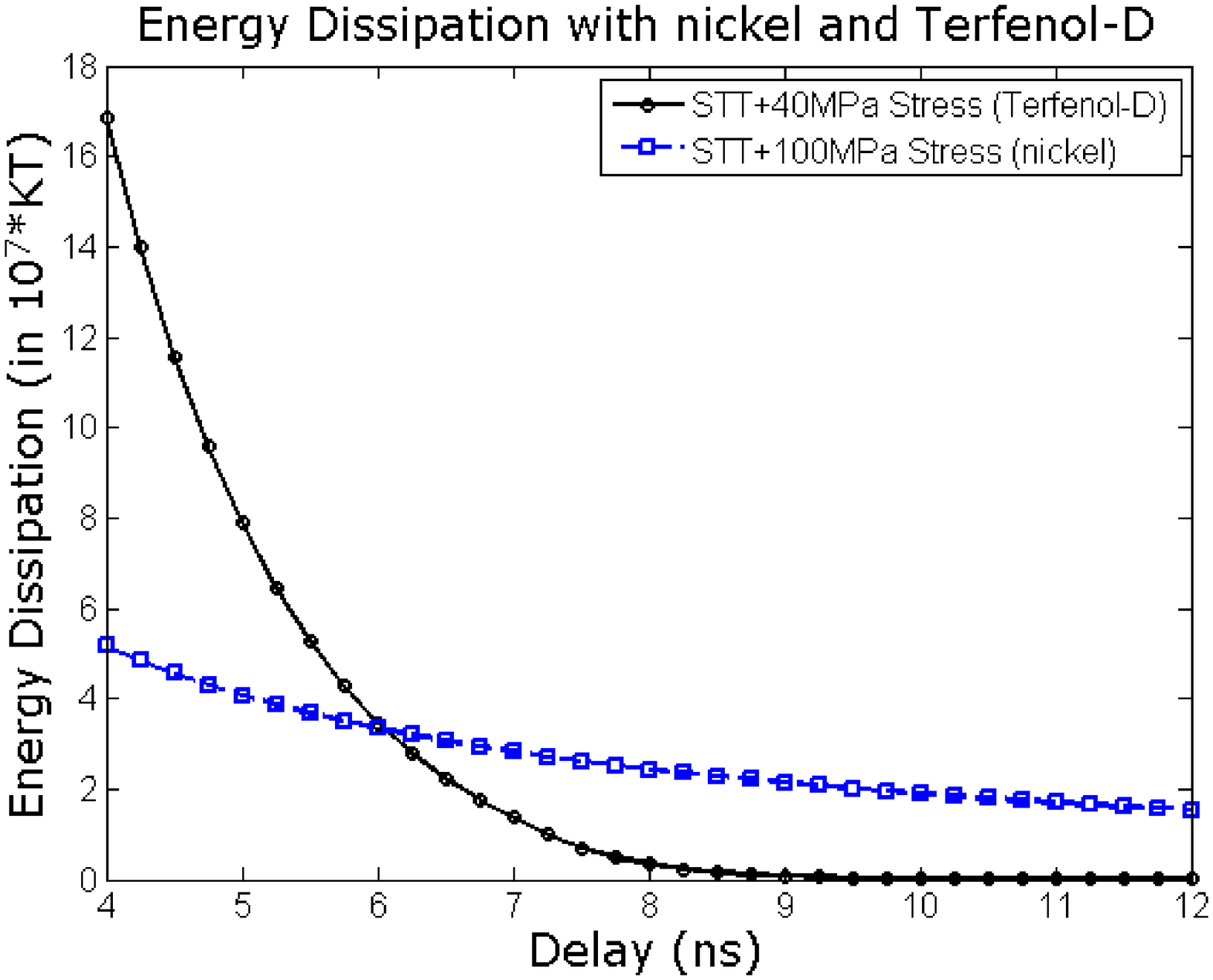}
\caption{\label{fig:comparison_energy_nickel_terfenolD} Energy dissipated in the MMS mode as a function of switching delay for Terfenol-D and nickel. The two curves are plotted for the maximum stress (40 MPa for Terfenol-D, and 100 MPa for nickel) that can be generated in the magnetostrictive layer assuming that the maximum possible strain in the PZT layer is 500 ppm. Note that MMS mode becomes more energy efficient in Terfenol-D for delays longer than 6 ns. Thus, for delays exceeding 6 ns, Terfenol-D will be the preferred material; otherwise, nickel will be preferred between the two.}
\end{figure}

Some additional simulation results and corresponding discussions are given in the 
Figures~\ref{fig:energy_stress_stt_mesh} -~\ref{fig:comparison_energy_nickel_terfenolD}.

\makeatletter 
\renewcommand\@biblabel[1]{[S#1]}
\makeatother

\clearpage
\pagebreak

\clearpage
\pagebreak